\documentstyle[epsfig,floats,prb,aps]{revtex}

\begin{document}
\title{\bf $ $ \\
Density of States and Energy Gap in Andreev Billiards}

\author{A. Lodder} 

\address{Faculteit Natuurkunde en Sterrenkunde, Vrije Universiteit,
         De Boelelaan 1081,\\ 1081 HV Amsterdam, The Netherlands}

\author{Yu. V. Nazarov}

\address{Faculteit der Technische Natuurkunde and DIMES, Technische Universiteit Delft,
         Lorentzweg 1, 2628 CJ Delft, The Netherlands}

\date{\today}
\maketitle

\begin{abstract}
\normalsize{
We present numerical results  for the local density
of states in semiclassical Andreev billiards.
We show that the energy gap near  the Fermi energy develops 
 in a  chaotic  billiard.  Using the same method no gap is
found in similar square and circular  billiards.}
\end{abstract}

% \twocolumn
\section{Introduction}
The density of states in a normal
metal in contact with a superconductor
is  affected  by the superconductor, as
a manifestation of the proximity effect or
%, more precisely, of the process (phenomenon) of
Andreev reflection.
In early days of superconductivity, this effect
has been observed in thin films of normal metal on a superconducting
substrate. \cite{Claeson}
It has been shown theoretically that for clean films
the spectrum of quasiparticle excitations remains gapless
\cite{DeGennes} at Fermi energy,
whereas a minigap develops for dirty films.\cite{McMillan}
The energy scale of this minigap is given by
$\hbar/t_N$, $t_N$ being a typical time spent by an electron
in the normal metal before it gets to the superconductor.\cite{McMillan}
 
Recent technological advances make it possible to
study the effect in more complicated geometries, in diffusive
metallic wires \cite{Gueron} and in a two-dimensional electron gas
where electron transport is almost ballistic \cite{Hartog}.
Recent theoretical developments \cite{frahm,altland}
suggested a new interpretation of old results.
\cite{DeGennes,McMillan} The existence of a minigap has been
related to {\it chaotic} character of the electron motion in the normal
part of the system. It makes no qualitative difference whether the
electron transport is diffusive, as in dirty films, or
chaotically ballistic, as in clean billiards \cite{berry}.
The absence of the gap in the deGennes spectrum follows from the fact
that a clean film is a specific case of a system with a separable geometry.
In such a system the  motion is not chaotic.
 
This interpretation is rather difficult to comprehend
in semiclassical terms. That was the motivation of the present
research. The problem is as follows.
The electron motion becomes truly chaotic (ergodic)
only for  trajectories that are very long in comparison with the
system size.
As we show in detail below, long electron trajectories correspond
to Andreev levels with energies very close to Fermi energy.
Therefore, in contrast to the interpretation in question,
one may argue that the presence of a minigap shows
that there are {\it no} long trajectories in the system.
Consequently, it may not be chaotic.
\begin{figure}[htb]
%\begin{minipage}{6.0cm}
%\centerline{\epsfig{figure=FigChCiSq.eps,height=5.0cm}}
\centerline{\epsfig{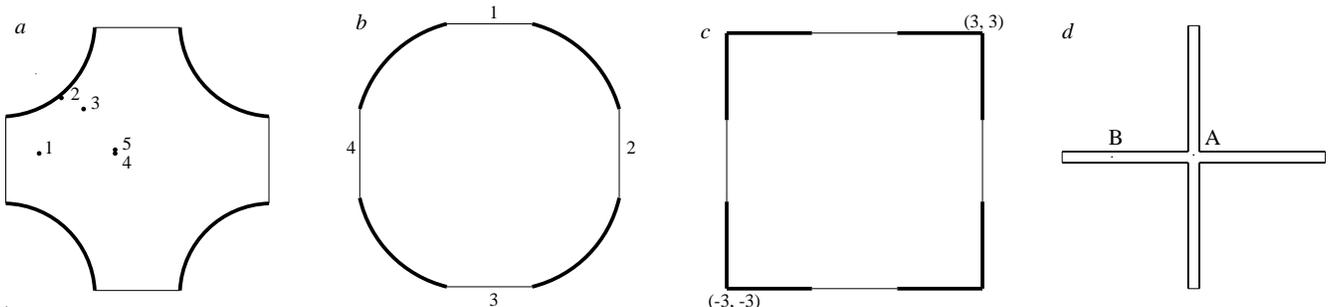}}
%\end{minipage}
\caption[]{Panels {\it a} to {\it c} show the three clean billiards investigated.
The specific
points looked at are indicated in the chaotic billiard. The
four normal/superconducting interfaces are labeled in the circular billiard.
Panel {\it d} shows the form of a quasi 1D diffusive system
investigated as well.}
\label{systems}
\end{figure}
 
To resolve this sophistry, we have calculated the density of states
for several Andreev billiards, chaotic and non-chaotic, depicted in
Fig. \ref{systems}$a$ to $c$.
Such billiards combine Andreev \cite{kosztin}and specular reflection
boundaries.\cite{berry} FIG. \ref{systems}$a$ shows the
form of the chaotic billiard
investigated. At the circular parts of the boundary specular reflection
occurs, while at the straight linear parts Andreev reflection takes
place. The outward concave shape of the circular parts make the system highly
chaotic. The special form chosen can be considered as representative
for any chaotic Andreev billiard. In order to show the marked
difference with related integrable Andreev billiards, calculations
are done also for circular and square boundaries. These systems
are depicted schematically in FIG. \ref{systems}$b$ and $c$. The
three systems differ only as far as the specularly reflecting
boundaries are concerned. Their Andreev reflecting boundaries are
identical. FIG. \ref{systems}$d$ shows the form of a quasi
one-dimensional diffusive system
investigated as well.

To find the density of Andreev states, we solve the equations for 
the quasiclassical Green's function 
along each classical trajectory. The solution depends explicitly on the
length $L$ of the trajectory considered, and gives a set of energy eigenvalues
$\simeq \hbar v_F/L$. 
Then, for each given point, we calculate numerically all possible classical trajectories
and sum up their contributions to the density of states.

To put our results shortly, for the chaotic system we do observe the formation
of a minigap near the Fermi level.
Long, truly chaotic trajectories appear to take an exponentially small fraction of
the phase volume and therefore do not contribute to the resulting density of states. 
The relevant equations are given in section \ref{theory}.
Results are discussed in section \ref{results}. Conclusions
are formulated in section \ref{conclusions}.

\section{The equation for local density of states}
\label{theory}

An expression for the local density of states $n(\epsilon, {\bf r})$
in the clean, ballistic systems to be considered, will be derived
by taking the imaginary part of the local Green's function
$G(\epsilon + i \delta, {\bf v}, {\bf r})$. The quasiclassical
Green's function $G(i\omega_{\rm n}, {\bf v}, {\bf r})$ to be
used will be a solution of the matrix equation \cite{eilenberger,ashida}

\begin{equation}
{\bf v}\cdot\nabla G + i [H,G] =0.
\label{greenvec}
\end{equation}
The velocity ${\bf v}$ is taken at the Fermi surface,
the matrix $H$ is given by

\begin{equation}
H=\left[\begin{array}{cc}
i\omega_{{\rm n}}   &   \Delta ({\bf r}) \\
-\Delta^{*}({\bf r}) &  - i\omega_{{\rm n}}
\end{array}\right],
\label{hmatrix}
\end{equation}
$\omega_{\rm n}$ are the Matsubara frequencies and $\Delta ({\bf r})$
is the superconducting gap function, to be taken constant in the
superconducting regions and zero in the central normal part of the system.
Eq. (\ref{greenvec}) can be solved analytically for any trajectory. By
accounting for all trajectories going through a given point the complete
local solution can be found. We will show this by first writing
Eq. (\ref{greenvec}) in the following from

\begin{equation} 
v_{{\rm F}}\frac{\partial}{\partial\ell} G + i [H,G] =0,
\label{greentrajec}
\end{equation} 
by which the ${\bf r}$ dependence is represented by the
length parameter $\ell$ along a trajectory.

Suppressing the $v_{\rm F}$ dependence, the general solution can be written in the form

\begin{equation}
G_{S}(i\omega_{{\rm n}}, \ell) = c_{1} A_{S} + c_{2} B_{S} 
e^{\frac{2\surd\omega_{\rm n}^{2} + \vert\Delta\vert^{2}}{v_{\rm F}}\ell} + c_{3} D_{S}
e^{-\frac{2\surd\omega_{\rm n}^{2} + \vert\Delta\vert^{2}}{v_{\rm F}}\ell},
\label{gensol}
\end{equation}
in which the matrices $A_{S}$ and $B_{S}$ are given by

\begin{equation}
A_{S}=\frac{-i}{\surd\omega_{\rm n}^{2} + \vert\Delta\vert^{2}}
\left[\begin{array}{cc}
i\omega_{{\rm n}}   &   \Delta  \\
-\Delta^{*} &  - i\omega_{{\rm n}}
\end{array}\right],
\label{As}
\end{equation}
and
 
\begin{equation}
B_{S}=\frac{1}{2(\surd\omega_{\rm n}^{2} + \vert\Delta\vert^{2})}
\left[\begin{array}{cc} 
-i\Delta^{*}  & \omega_{\rm n}+\surd(\omega_{\rm n}^{2} + \vert\Delta\vert^{2}) \\
-\frac{\Delta^{*}}{\Delta}(\omega_{\rm n} -
\surd(\omega_{\rm n}^{2} + \vert\Delta\vert^{2})& i\Delta^{*}
\end{array}\right], 
\label{Bs} 
\end{equation} 
while $D_{S}= B_{S}^{\dagger}$. This solution is most easily obtained in two
steps. First Eq. (\ref{greentrajec}) is solved for the normal system,
taking $\Delta ({\bf r}) = 0$ in the matrix $H$. One finds

\begin{equation}
G_{N}(i\omega_{{\rm n}}, \ell) = c_{4} A_{N} + c_{5} B_{N}
e^{\frac{2\omega_{\rm n}\ell}{v_{\rm F}}} + c_{6} D_{N}
e^{-\frac{2\omega_{\rm n}\ell}{v_{\rm F}}},
\label{norsol}
\end{equation}
with matrices

\begin{equation} 
A_{N}=\left[\begin{array}{cc}
1 & 0 \\
0 & -1
\end{array}\right],\hspace{1mm} {\rm and}\hspace{1mm} B_{N}= \left[\begin{array}{cc}
0 & 1 \\
0 & 0
\end{array}\right],
\label{An} 
\end{equation}
%\begin{equation}
%B_{N}=\left[\begin{array}{cc}
%0 & 1 \\
%0 & 0
%\end{array}\right], 
%\label{Bn}  
%\end{equation}
while $D_{N}= B_{N}^{\dagger}$. Since Eq. (\ref{greentrajec}) is a
homogeneous equation, the solution (\ref{norsol}) is complete apart from an
overall constant, to be determined by the
requirement, that the matrix $A_{N}$ times that constant is the solution of
the original equation for the Green's function of a bulk system, still having
a delta function at the right hand side. This will merely lead to
the proper normalization \cite{eilenberger} of the expression for
the density of states to be derived below.

In the second step the full matrix $H$ is diagonalized by
the unitary matrix

\begin{equation}
U=\frac{1}{\surd 2}\left[\begin{array}{cc}
\surd(1 + \frac{\omega_{\rm n}}{\surd(\omega_{\rm n}^{2} + \vert\Delta\vert^{2}})
&  \frac{i\Delta}{\vert\Delta\vert}\surd(1 -
\frac{\omega_{\rm n}}{\surd(\omega_{\rm n}^{2} + \vert\Delta\vert^{2}}) \\
\frac{i\vert\Delta\vert}{\Delta}\surd(1 - 
\frac{\omega_{\rm n}}{\surd(\omega_{\rm n}^{2} + \vert\Delta\vert^{2}}) &
\surd(1 + \frac{\omega_{\rm n}}{\surd(\omega_{\rm n}^{2} + \vert\Delta\vert^{2}})
\end{array}\right].
\label{Usn}
\end{equation}
The correspondingly transformed Eq. (\ref{greentrajec}) has the same form
as the equation for the normal system, the decay length being replaced
by $v_{\rm F}/\surd(\omega_{\rm n}^{2} + \vert\Delta\vert^{2})$.
Consequently, the full solution
Eq. (\ref{gensol}) of Eq. (\ref{greentrajec}) now can be obtained by
the unitary transformation $U G_{N} U^{\dagger}$ of the matrix given 
by Eq. (\ref{norsol}), and substituting the proper decay length.

The solution of Eq. (\ref{greentrajec}) for a trajectory in
inhomogeneous systems as depicted in FIG. \ref{systems} is obtained as follows.
Consider an arbitrary trajectory, hitting some point at one of the
superconducting/normal interfaces, at which point an Andreev
reflection takes place, and follow the trajectory inside the normal
region by accounting for all specular reflections at the boundaries, until
it hits a superconducting/normal interface again. If the length
parameter ${\ell}$ is taken 0 at the initial hitting point and equal
to $L$ at the second hit, $L$ standing for the total length of
the trajectory, the form of the solutions in the different
regions is clear from the Eqs. (\ref{gensol}) and (\ref{norsol}).
For ${\ell} \leq 0$ and $\ell \geq L$ the solution (\ref{gensol})
is to be used, Since the dimensions of the superconducting regions
are supposed to be large, behaving effectively as bulk superconductors,
the coefficient of the matrix $A_{S}$ can be taken to be equal to
1. Further, for $\ell < 0$ the coefficient $c_{3}$ has to be
put 0, while for $\ell > L$ the $B_{S}$ term blows up and
the corresponding coefficient has to be put 0. The normal region
is supposed to have mesoscopic dimensions, and the full solution 
$G_{N}(i\omega_{{\rm n}}, \ell)$ has to be used. The equation
(\ref{greentrajec}) being a first order differential equation,
the only requirement is continuity at the interfaces. By that
one finds for $c_{4}$ the following expression

\begin{equation} 
c_{4} = \frac{\omega_{\rm n}}{\surd\omega_{\rm n}^{2} + \vert\Delta\vert^{2}}
+ \frac{\vert\Delta\vert^{2} \tanh \frac{\omega_{\rm n}L}{v_{\rm F}}}
{\omega_{\rm n}^{2} + \vert\Delta\vert^{2}
 + \omega_{\rm n}\surd(\omega_{\rm n}^{2} + \vert\Delta\vert^{2})
\tanh \frac{\omega_{\rm n}L}{v_{\rm F}}}.
\label{cvier}
\end{equation}

\begin{figure}[b]
%\begin{minipage}{16.0cm} 
\centerline{\epsfig{figure=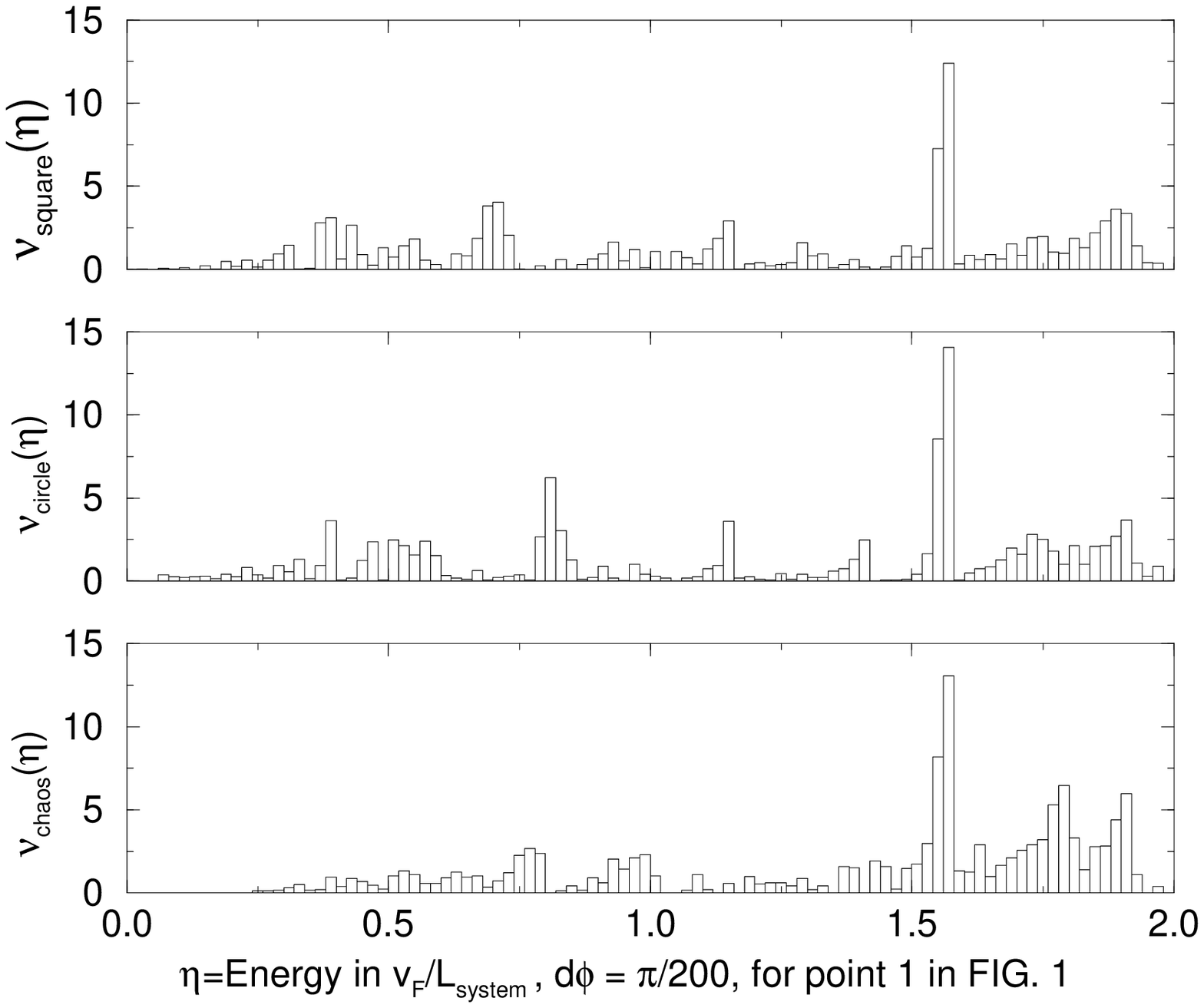,height=9.0cm,width=9.0cm}}
%\end{minipage}    
\caption[]{From top to bottom the local density of states for the square,
circular and chaotic billiards respectively, at point 1 in FIG. \ref{systems},
for the ${\phi}$-grid indicated along the horizontal axis. 
}
\label{f100}
\end{figure}

Now everything is ready for the local density of states in
the normal region. First of all, since we are after studying
the development of a gap just above the Fermi energy and
of a width much smaller than $\vert\Delta\vert$, it is sufficient
to focus on the coefficient $c_{4}$ in the limit 
$\vert\omega_{\rm n}\vert \ll \vert\Delta\vert$, so on
 
\begin{equation}
c_{4} = \tanh \frac{\omega_{\rm n}L}{v_{\rm F}}.
\label{limc4}
\end{equation}

Secondly, only the left upper matrix element of $G_{N}(i\omega_{{\rm n}}, \ell)$
is required for the density of states, so only the $c_{4}$ term
in Eq. (\ref{norsol}) contributes. After the substitution
$i\omega_{\rm n} \to \epsilon + i\delta$, one finds the contribution
of a trajectory of length $L$ through a chosen point, 
to the density of states at that point, to
be proportional to

\begin{equation}
 \lim_{\delta \to 0}
\Im i\tanh\frac{(\epsilon + i \delta) L}{iv_{\rm F}}
= \lim_{\delta \to 0}\Im
\tan\frac{(\epsilon + i \delta) L}{v_{\rm F}}=
\pi\sum_{n} \delta (\frac{\epsilon L}{v_{\rm F}} -
(n + \mbox{$\frac{1}{2}$})\pi),
\label{dosLtan}
\end{equation}
in which the summation runs over integer {\it n} values.
For a given point in the interior region
of a 2D system all trajectories through
that point have to be taken into account. This leads to the following expression for
the dimensionless local density of states $\nu (\epsilon, {\bf r})$,
defined by
  
\begin{equation}
\nu (\epsilon, {\bf r}) \equiv \frac{n (\epsilon, {\bf r})}{n_{N}}=
\int_{0}^{\pi} {\rm d}\phi \sum_{n} \delta (\frac{\epsilon L(\phi)}{v_{\rm F}} -
(n + \mbox{$\frac{1}{2}$})\pi),
\label{dosnuE}
\end{equation}
in which $n_{N}$ is the constant density of states of a 2D normal
system, and $\phi$ is the slope angle of a trajectory with length $L(\phi)$.
For presenting the results it is most convenient to shift to
the dimensionless energy variable $\eta \equiv \epsilon L_{\rm system}/v_{\rm F}$,
in which $L_{\rm system}$ stands for the length dimension of the system.
By that, and denoting the relative density of states, now depending
on the variable $\eta$, by the same symbol $\nu$, we end up at the expression

\begin{equation} 
\nu (\eta, {\bf r}) = \int_{0}^{\pi} {\rm d}\phi \sum_{n} \delta (\eta 
\frac{L(\phi)}{L_{\rm system}} -
(n + \mbox{$\frac{1}{2}$})\pi).
\label{doseta}
\end{equation} 
Note that in this final expression the length of a trajectory is present in a relative
way, and has become dimensionless as well.

\section{Results and discussion}
\label{results}

Because Eq. (\ref{doseta}) contains the relative quantity $L(\phi)/L_{\rm system}$
only, the physical size of the system does not enter. However, in actual
calculations a choice has to be made. We have chosen
 $L_{\rm system}$
to be  equal to 6. In all systems depicted in FIG. \ref{systems}
the origin lies in the center of the billiard. The corners of the square
billiard then lie at $(\pm3, \pm3)$. The
length of the S/N interfaces is chosen to be equal to 2. The specific
points {\it 1} to {\it 5} to be looked at have the coordinates ${\it 1}=(-2.2, 0.1)$,
${\it 2}=(-1.7, 1.4)$, ${\it 3} = (-1.2, 1.1)$, ${\it 4} = (-0.5, 0.1)$ and
${\it 5} = (-0.5, 0.2)$.

\begin{figure}[htb]
%\begin{minipage}{16.0cm}
\centerline{\epsfig{figure=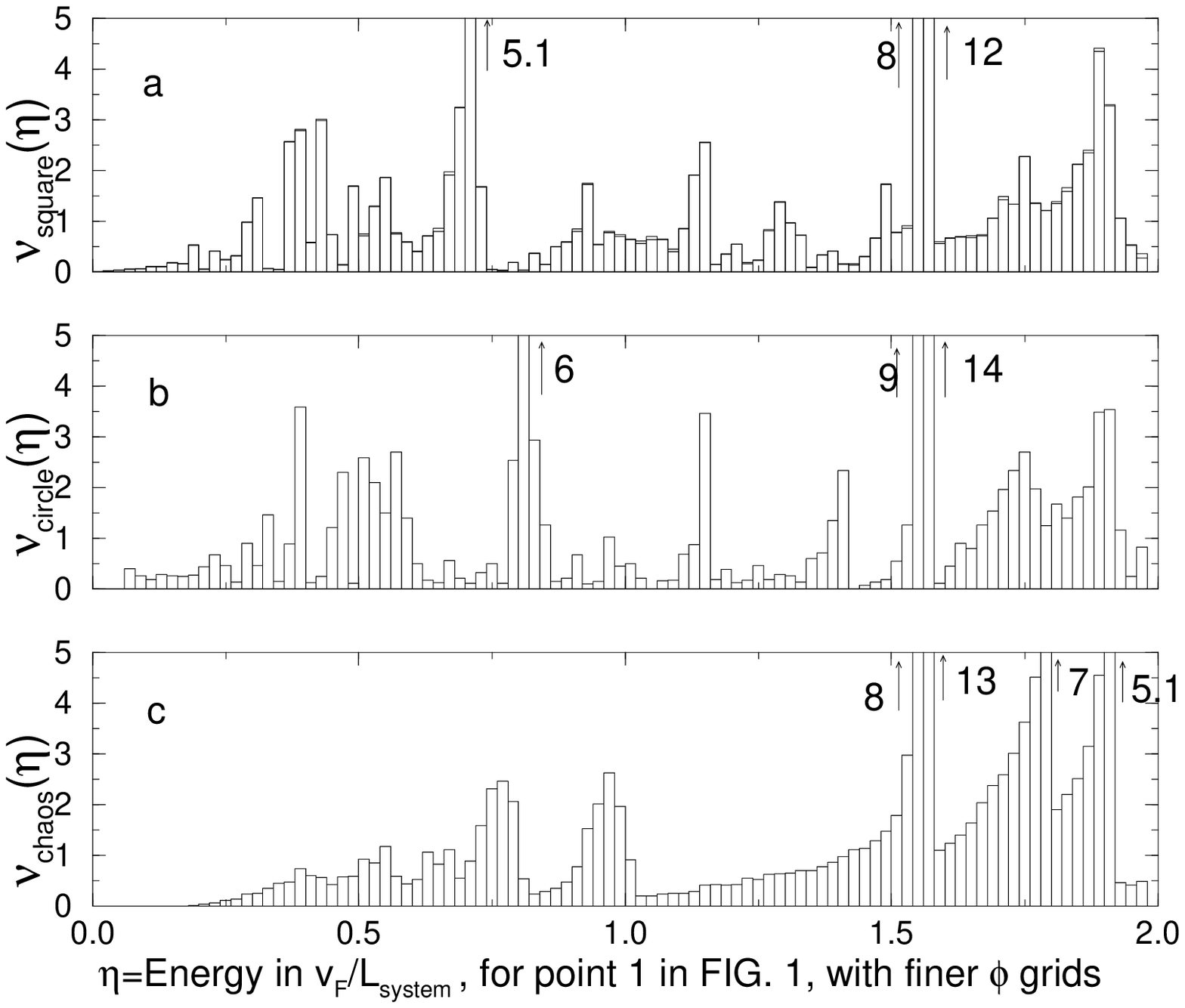,height=9.0cm,width=9.0cm}}
%\end{minipage}
\caption[]{The same as in FIG. \ref{f100}, but for both a ten and hundred times finer
${\phi}$-grid for the square billiard, and for a hundred times finer
grid for the circular and chaotic billiards.
}
\label{f1000}
\end{figure}

In FIG. \ref{f100} the density of states is shown for point {\it 1}
in the three billiards under consideration, for a ${\phi}$-grid with $d\phi = \pi/200$.
For comparison the results according to finer grids are shown in FIG.
\ref{f1000}. In this figure the scale along the vertical axis has
been adapted, by which the high peaks are not displayed fully, but
the average structure comes out more clearly. The peak heights have been
given explicitly.
For the chaotic billiard, both figures show a gap at the Fermi energy,
which corresponds with ${\eta} = 0$.
The missing states in FIG. \ref{f100} near ${\eta} = 0$
for the square billiard are apparently due to the grid, because no gap is seen
if ten and hundred times finer meshes are used. The finer mesh leads to a histogram,
which is hardly distinguishable from a $d\phi = \pi/2000$ picture. But
because of security all other histograms to be displayed are
obtained using a $d\phi = \pi/20000$ mesh. The density of states
${\nu}_{\rm square} (\eta)$ is quite similar for the different
points, and no particular features show up, so we further
concentrate on the circular and chaotic billiards.

\begin{figure}[tb]
%\begin{minipage}{16.0cm}
\centerline{\epsfig{figure=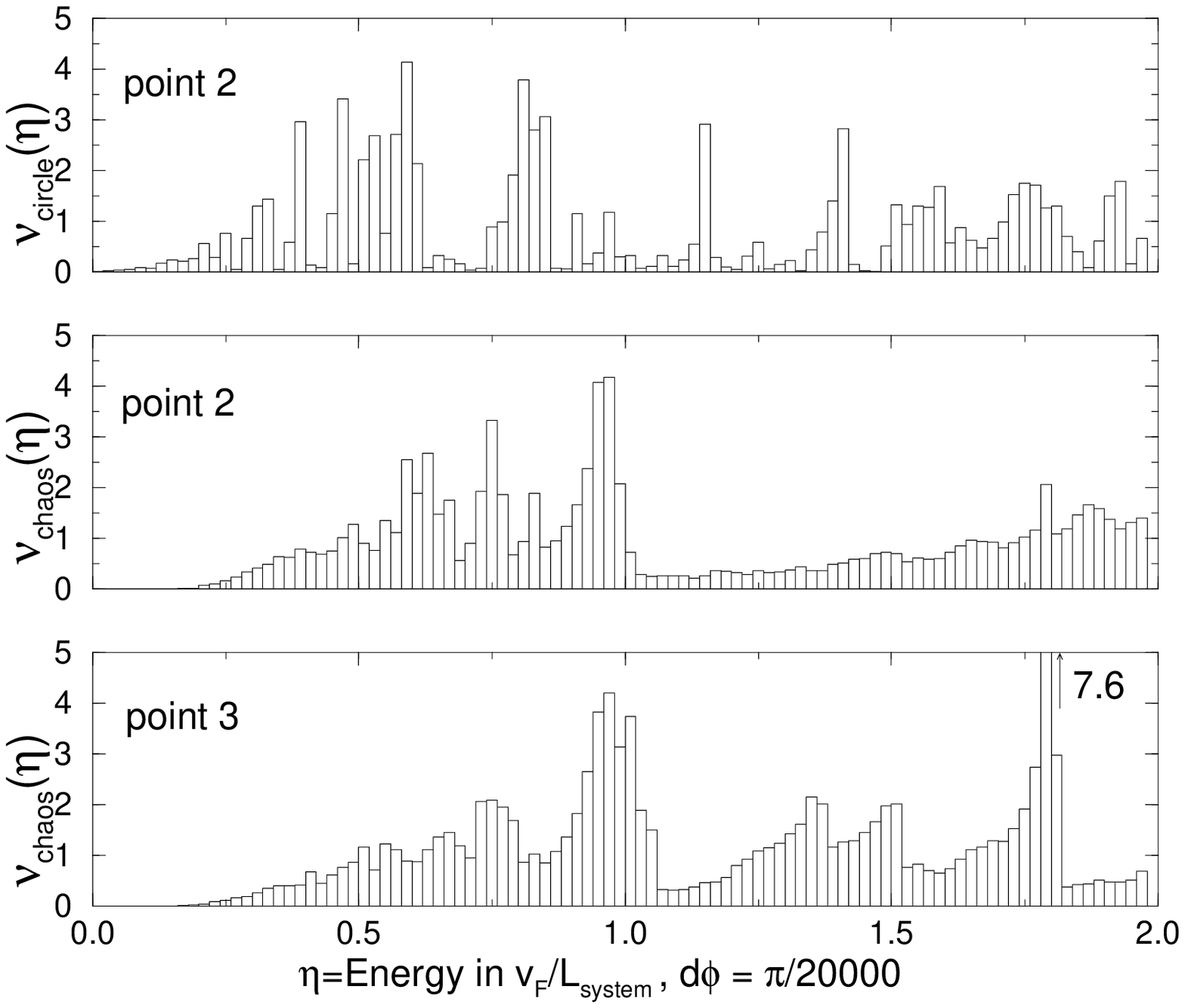,height=9.0cm,width=9.0cm}}
%\end{minipage}
\caption[]{The local density of states for the circular billiard at point 2,
and for the chaotic billiard at the points 2 and 3 in FIG. \ref{systems}.
}
\label{f1001}
\end{figure}

\begin{figure}[tb]
%\begin{minipage}{16.0cm}
\centerline{\epsfig{figure=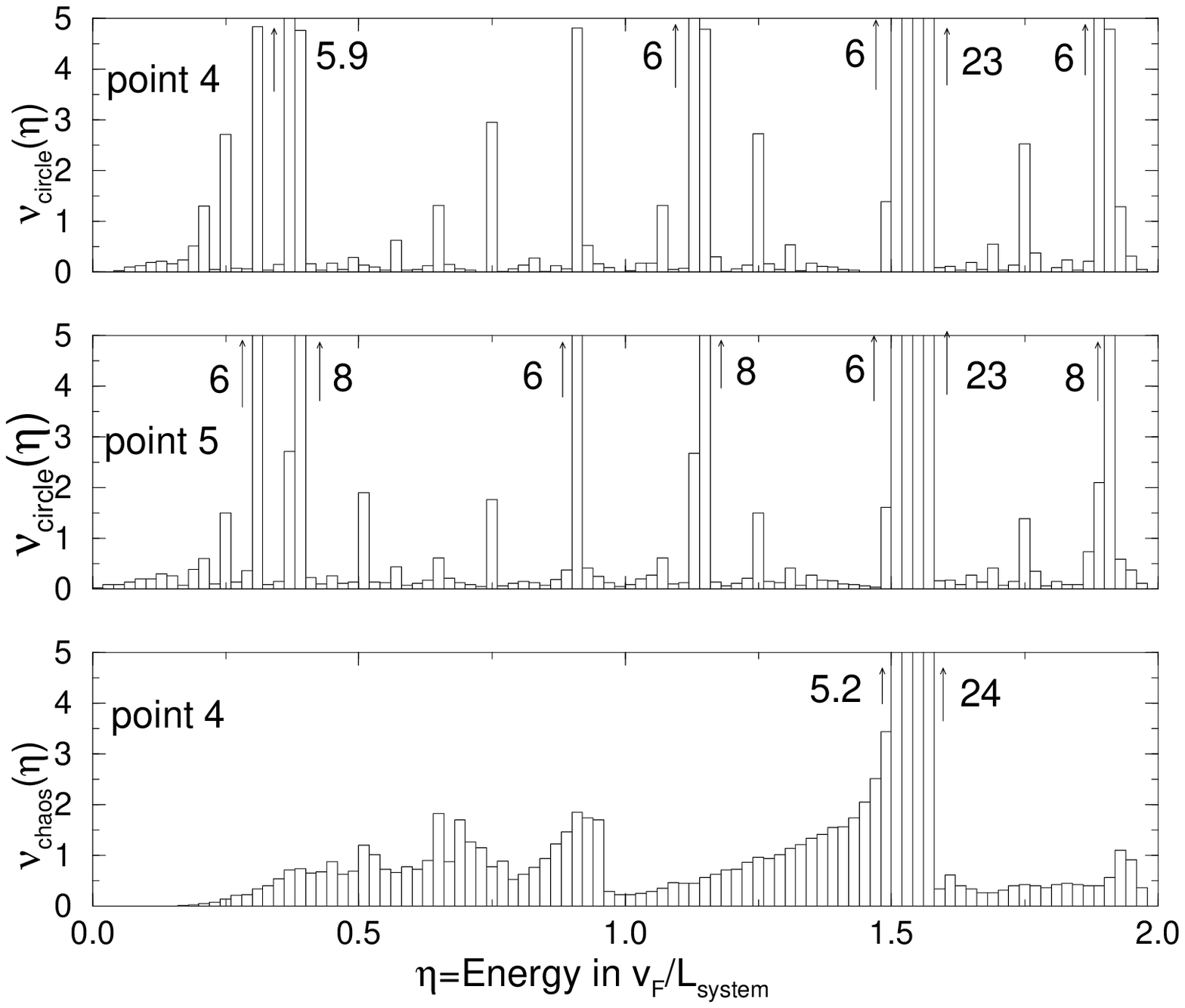,height=9.0cm,width=9.0cm}}
%\end{minipage}
\caption[]{The local density of states for the circular billiard
at the points 4 and 5, and for the chaotic billiard at point 4.
}
\label{fCh123}
\end{figure}
 
The gap for the circular billiard is certainly
typical for the point chosen. This becomes clear if one
realizes, that states near ${\eta} = 0$ are due to trajectories, which are long
compared to the system size. For the circular billiard such long trajectories 
contribute only if the line through a
chosen point and the center of the circle hits a normal,
specularly reflecting boundary. This requirement is not fulfilled
for point {\it 1}. At the top of FIG. \ref{f1001} a similar plot is given, but
for point {\it 2}, for which such trajectories certainly contribute.
Now no gap is seen for the circular billiard. The gap for the chaotic billiard
is manifestly present in the middle of this figure, for point {\it 2}, and at the
bottom, for point {\it 3}. The gap in the histogram for the circular billiard
depends critically on the precise location of the point. This is shown
in FIG. \ref{fCh123}, displaying results for the points {\it 4} and
{\it 5}. While point {\it 4} does not support long trajectories,
point {\it 5} does. For the latter point the gap is closed,
although it lies very near to point {\it 4}. We show the histogram
for the chaotic billiard at one of these points only, because not much 
difference is seen for this latter system.

For the chaotic system the gap is present for all points. It appears to
be an intrinsic property of this system. Long trajectories are
rare, irrespective of the point considered. While for the square
and circular billiards trajectories with over 2000 times the
system size are easily found, for the chaotic billiard it is hard to
find trajectories longer than 20 times the system size. This is 
illustrated by TABLE \ref{tablephi}.
The data given are obtained
as follows. First, for fifty points the longest trajectory was
calculated, using for each point a $\phi$ grid of $d\phi = \pi/2000$.
The longest of the 100000 trajectories considered this way was
found for point (-1.2, 0.9) at the angle given in the first
line of the table in column 1. In addition to its length of 71.8 its relative length is 
given. In the last three columns the number of specular reflections and the
labels of the S/N interfaces of both ends of
the trajectory, called exits, are given. The meaning of these
labels is shown for the circular billiard, in the middle of
FIG. \ref{systems}. After that the angle was specified finer and
finer, using 10 angle values on both sides of the angle considered.
Progressively the angle giving longest of the 21 lengths calculated that way
was picked out for further subdivision. At the end, at one angle
a length of 19.08 times the system size was found, but then the borderline
of our (double) precision was reached. The sensitivity
to the initial condition is illustrated by the results
on the 5th and 8th line from the bottom, because they hold for
the same angle. This angle was generated in a slightly different way in the
subsequent subdivision. We conclude that long trajectories are rare,
although, theoretically they are supposed to exist. For example,
comparing the trajectories illustrated by the third and fourth
line, there must be an angle $\phi_0$ in between, at which 
the shift occurs from the first exit to the second one. This critical angle would
support an infinitely long trajectory.

%Insert
To give some qualitative estimations, we consider the contribution of the trajectories
nearing that critical angle $\phi_0$. The length at $\phi \rightarrow \phi_0$
can be estimated
as $L(\phi) \simeq - L_{\rm system} \log_2(\vert\phi-\phi_0\vert)$. This estimation
follows from the fact that
each bounce between concave boundaries approximately doubles the deviation angle.
Using the relation (\ref{doseta}) we estimate $\nu \simeq 2^{-\frac{\pi}{2\eta}}$.
This shows that the density of states
is exponentially suppressed at small $\eta$. With
our numerical method, having a finite grid $\delta \phi$, we can only access 
energies $\eta \ge - \log_2(\delta \phi)$. The states
with smaller energies are not seen. The good convergence of our numerical
data, even at relatively small $\eta$,
proves that the gap develops rather quickly, giving rise to an
abrupt change of the density of states.

It is interesting to note that quantummechanical effects can also be 
estimated in this way.
Due to diffraction of the electron waves in  a billiard geometry, the best
angle resolution is limited by
$\delta \phi_{\hbar} = 1/\sqrt{k_{\rm F} L_{\rm system}}$, $k_{\rm F}$
being the electron wave vector at Fermi energy.
This implies that no Andreev states exists below
the threshold energy $\eta \simeq 1/\log_2(k_{\rm F} L_{\rm system})$.
%End of insert
\begin{table}[t] 
\caption{Illustration of the fact, that long trajectories are rare in the chaotic
billiard. From top to bottom increasing lengths $L(\phi)$ are found for an increasingly
precisely chosen angle $\phi$. The data are for a point somewhat below point {\it 3},
with coordinates (-1.2, 0.9).}
\begin{tabular}{cccccc}
$\phi$ in radians&$L(\phi)$&$L(\phi)/L_{\rm system}$&No. reflections&Exit1&Exit2 \\
\hline   
 -0.92677083000000  &  71.8 &  11.96 &  19  &   1 & 3 \\
&&&&& \\
 -0.92677083600000  &  75.5  & 12.59 &  21  &   1 & 1 \\
 -0.92677083700000  &  92.4 &  15.40&   25 &    1&  1 \\
 -0.92677083800000  &  70.8 &  11.80 &  20  &   1&  2 \\
&&&&& \\
 -0.92677083690000  &  82.5&   13.76 &  23 &    1&  4 \\
 -0.92677083700000  &  92.4 &  15.40 &  25  &   1&  1 \\
 -0.92677083710000  &  77.3 &  12.88 &  21 &    1&  4 \\
&&&&& \\
 -0.92677083700000  &  92.4 &  15.40  & 25   &  1 & 1 \\
 -0.92677083701000  &  96.5 &  16.09 &  27  &   1 & 1 \\
 -0.92677083702000  &  77.8 &  12.97 &  22  &   1 & 4 \\
&&&&& \\
 -0.92677083700000  &  92.4 &  15.40 &  25 &    1&  1 \\
 -0.92677083700100 &  110.3  & 18.39 &  30 &    1&  3 \\
 -0.92677083700200 &   84.2  & 14.04 &  24 &    1&  1 \\
&&&&& \\
 -0.92677083700090 &   88.9  & 14.82 &  25 &    1&  4 \\
 -0.92677083700100 &  107.2  & 17.86 &  30 &    1&  4 \\
 -0.92677083700110 &   91.0  & 15.16 &  25 &    1&  3 \\
&&&&& \\
 -0.92677083700096 &   98.0  & 16.33 &  27 &    1&  4 \\
 -0.92677083700097 &  114.5  & 19.08 &  32 &    1&  2 \\
 -0.92677083700098 &   94.2  & 15.69 &  26 &    1&  2 \\
\hline
\end{tabular}
\label{tablephi}
\end{table}

Although it is not the primary aim of the present paper,
it is interesting to discuss the structure seen in the histograms.
The peaks are certainly due to the fact, that special classes of
trajectories contribute more than an average trajectory. Considering
the $L({\phi})$ dependence of the local density of states
expression (\ref {doseta}), for a given point and in an arbitrary direction
this function will contain a term linear in $\phi$. But there are
exceptions. The length $L({\phi})$ for trajectories through the points {\it 1},
{\it 4} and {\it 5} in FIG. \ref{systems} behaves quadratically in 
$\phi$ around ${\phi}=0$,
while for the latter two points in addition a quadratic behaviour in $\phi- {\pi}/2$
around ${\phi}={\pi}/2$ holds. Then the argument of the delta function in
Eq. (\ref {doseta}) behaves quadratically in $\phi$ around ${\phi}=0$, giving rise to
a {\it square root singularity} in the density of states. This
singularity produces  a series of equidistant peaks in the histogram with peak energies
counted by the integer {\it n}. At ${\phi}=0$
the length $L({\phi})$ is equal to $L_{\rm system}$, so the lowest
peak, for ${\it n} = 0$, is expected to be seen at $\eta = {\pi}/2 = 1.57$.
This peak is easily recognized in the FIGs. \ref{f100}, \ref{f1000}
and \ref{fCh123}.
Another type of trajectory in a direction ${\phi}_{0}$, around which $L({\phi})$ behaves
quadratically in $\phi- {\phi}_{0}$, is, for example, the one
through point {\it 3} in FIG. \ref{systems} with ${\phi}_{0} \approx {\pi}/4$,
hitting the exits {\it 1} and {\it 4}.
Since the line from the origin to point {\it 3} has a slope that is
slightly smaller than -1, ${\phi}_{0}$ is not equal to ${\pi}/4$.
The corresponding $L({\phi}_{0})=5.23$, leading to a
peak at $\eta =1.8$, which is clearly present in the
lower panel of FIG \ref{f1001}. The
corresponding peak for point {\it 2} lies too much to the
right to be seen, namely at $\eta =2.3$. We just point to another
class of trajectories giving rise to quadratic behaviour, namely,
for example, the one through point {\it 3} and perpendicular to any
outward concave circle. The corresponding line connects point {\it 3} and 
the center of the circle. Although $L({\phi})$ behaves quadratic as far
as the contribution towards the circle is concerned only, possible
linear contributions from the backward part of the trajectory
will lead to a shift of the extreme ${\phi}_{0}$. We calculated the
corresponding lengths, and, just as an illustration, we mention
the peaks for a few points. The trajectory through point {\it 1}
perpendicular to the circle centered at (3, 3) leads to a peak
at 0.97, which is clearly seen in the lower panels of FIGs. \ref{f100}
and \ref{f1000}. The trajectories through point {\it 2} perpendicular
to the circles centered at (3, 3) and (3, -3) lead to peaks at
0.97 and 0.49 in the middle panel of FIG. \ref{f1001}.
 
\begin{figure}[tb]
%\begin{minipage}{11.0cm}
\centerline{\epsfig{figure=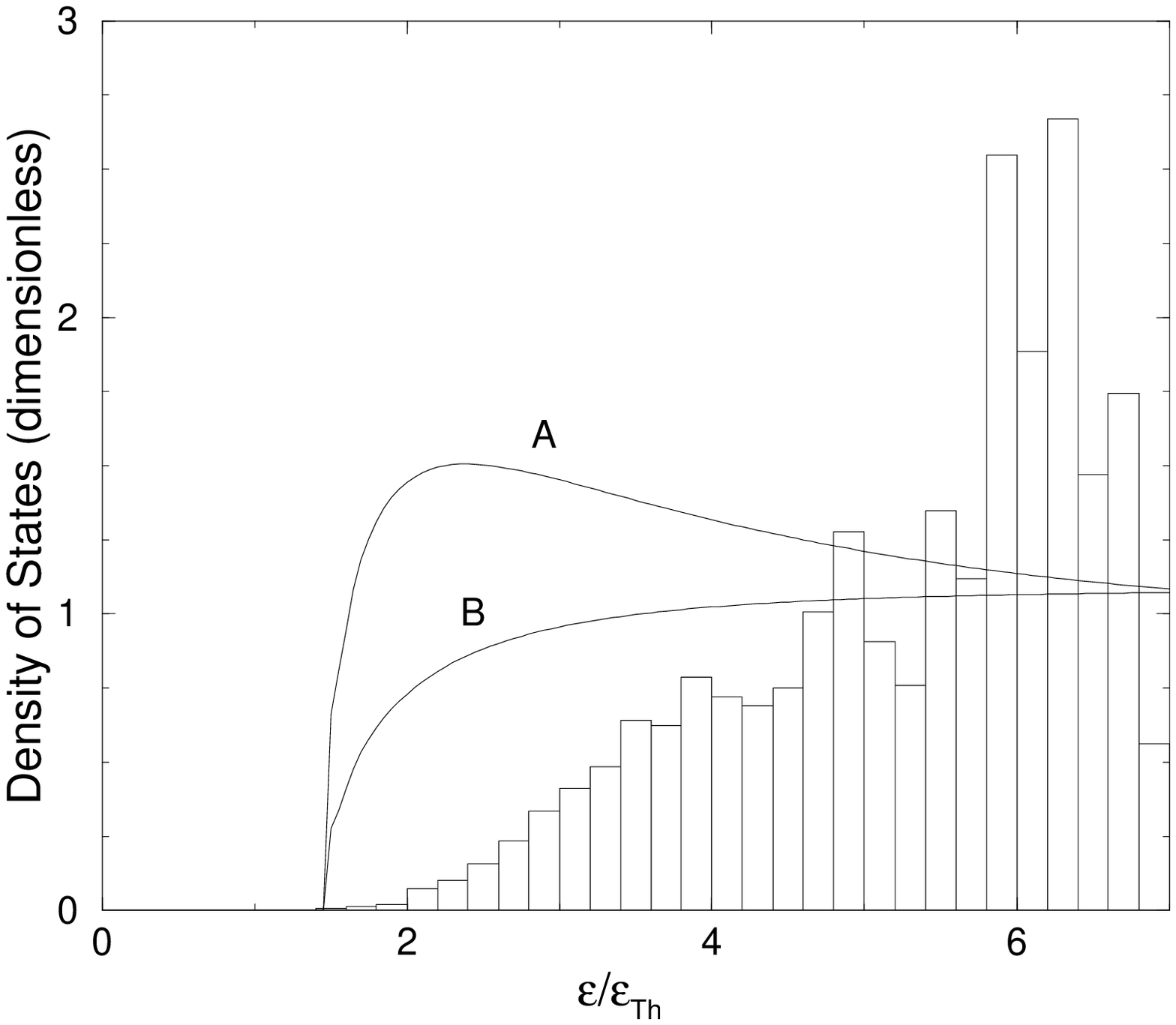,height=6.0cm}}
%\end{minipage}
\caption[]{The local density of states at the points A and B in the
1D dirty system depicted in FIG. \ref{systems}{\it d}, compared with the histogram for
point 2 in the 2D clean chaotic system depicted in
FIG. \ref{systems}{\it a}.}
\label{fdirty}
\end{figure}

Finally we want to point at even another type of extremal trajectory,
namely a trajectory through a point and touching a circle. Consider
the trajectory through point {\it 2}, in upward direction touching
the circle  centered at (-3, 3). 
The length has a minimum value
at the touching angle ${\phi}_{0}$, but the $\phi$ dependence remains
linear. Still an effect can be expected, because the coefficient
of $\Delta \phi \equiv \phi- {\phi}_{0}$ for positive $\Delta \phi$ can be 
different from the coefficient for negative $\Delta \phi$.
This leads to a {\it step}
in the evaluation of the delta function in the expression for
the local density of states, because $\delta (b \Delta \phi) =
\delta (\Delta \phi)/|b|$. In analyzing the different touching
trajectories for the different points in the chaotic billiard,
not all possible steps are clearly
recognized, and others correspond to  values of $\eta$, which
lie too high to be seen. We just mention the trajectory through
point {\it 2} touching the circle around (3, 3), which is expected
to give a step at $\eta =0.77$, and the trajectory through
point {\it 3} touching the circle around (-3, -3), corresponding
to a step at $\eta =1.35$. Both steps can be recognized in the
middle and lower panel respectively of FIG \ref{f1001}. 

We conclude by comparing the results for some point in the clean 2D chaotic
billiard depicted by FIG. \ref{systems}{\it a}
with the local density of states at a central point A and a point B
closer to the S/N interface in the
{\it dirty} 1D Andreev billiard depicted in
FIG. \ref{systems}{\it d}. The results for the dirty system
are obtained by numerical integration of the Usadel equations.
\cite {usadel} The curves A and B in FIG. \ref{fdirty} 
show the dimensionless density of states for the two
points in the dirty 1D system, while the histogram holds for
point 2 of the clean system and reproduces the
lower energy part of the middle panel of FIG. \ref{f1001}.
Mind, that the scale of the gap energy $E_g$ is different
in the ballistic and diffusive case: for a ballistic system 
$E_g \simeq \hbar v_{\rm F}/L_{\rm system}$,
whereas for a diffusive system it is strongly reduced,
$E_g \simeq \hbar v_{\rm F}/\sqrt{lL_{\rm system}}$,
$l$ being the mean free path. In order to compare results, the curves
and the histogram are rescaled to a common $E_g$.

Despite the gap occurs in both cases, it is seen that the behavior of the
density of states is quite different for diffusive and ballistic cases.
Consequently, this behavior is not universal and depends on details of the geometry
and the scattering within the billiard. This fact strongly reduces the applicability
of Random Matrix Theory methods, which are based on the assumption of universality
of chaotic behavior. We note that the absence of universality can be understood
from the fact
that long and truly chaotic trajectories do not contribute to the density of states.
Therefore it is determined by non-ergodic, non-universal trajectories.  

\section{Conclusions and prospects}
\label{conclusions}
%Insert
Our results prove that a gap is formed in the density of states of a chaotic 
Andreev billiard even in the semiclassical limit. This is despite the fact that
in the semiclassical limit Andreev states could have a very small energy being generated
by very long trajectories. It turns out that the density of states of a chaotic billiard
exhibits an abrupt drop at energies several times smaller than $v_{\rm F}
/L_{\rm system}$. Below
this energy, the density of states is exponentially suppressed. We believe that quantum
mechanical effects will lead to complete exhausting of the density of states
at energies below $v_{\rm F}/(L_{\rm system} \log (k_{\rm F}
L_{\rm system}))$. Comparison of the density
of states in the billiard and in the diffusive system clearly shows the absence of
even approximate universality.
 
For comparison a square and a circular Andreev billiard have been considered. 
In the square system long trajectories are always present, and no gap develops. 
Interestingly, although in roughly 70\% of the volume of the circular system
no gap is visible  in the local density of states, in the remaining
part of the volume there is also a gap developing. This possibly can be explained
by the fact that the billiard is not ideally circular and may be slightly chaotic.

All three billiards exhibit geometrically induced features in the density of states.
Trajectories that traverse a billiard perpendicular to the
superconducting boundaries, without reflection,
generate a sequence of equidistant square root singularities.
Trajectories that touch concave boundaries lead to steps in the density of states.
%End of insert

%\acknowledgments
%  One of us (AFT) wants to acknowledge

%\appendix
%\section{}
%\label{appendixA}

\end{document}